# ULTRA-WIDEBAND TAPERED TRANSDUCERS IN THIN-FILM LITHIUM NIOBATE ON SILICON CARBIDE

*Jack Kramer, Tzu-Hsuan Hsu, Joshua Campbell, and Ruochen Lu*
University of Texas at Austin, TX, USA

## ABSTRACT

Acoustic devices offer significant advantages in size and loss, making them ubiquitous for mobile radio frequency signal processing. However, the usable bandwidth is often limited to the achievable electromechanical coupling, setting a hard limit using typical transducer designs. In this work, we present an ultra-wideband transducer design utilizing a tapered electrode configuration to overcome this limitation. The design is realized on a lithium niobate (LN) on silicon carbide platform, utilizing a combination of first and higher order shear-horizontal modes to generate the ultra-wideband response. The implementation shows a fractional bandwidth (FBW) of 55% at 2.23 GHz with an associated insertion loss (IL) of 26 dB for the measured 50 Ω case. Upon improved impedance matching, this performance could be improved to 79% FBW and an IL of 16.5 dB. Upon further development, this ultra-wideband design could be reasonably scaled towards improved FBW and IL trade off to enable improved usability for cases where bandwidth should be prioritized.

## KEYWORDS

Acoustic Devices, Lithium Niobate, Piezoelectric, Surface Acoustic Wave

## INTRODUCTION

Piezoelectric acoustic devices are widely used in modern mobile communication systems, owing their success to low loss and compact form factors compared to electromagnetic counterparts. These advantages have made platforms such as surface acoustic wave (SAW) and bulk acoustic wave (BAW) ubiquitous in mobile devices towards the realization of components such as filters and delay lines [1], [2]. SAW in particular benefits from simpler fabrication processes, making it dominant for lower frequency applications where critical features can be patterned consistently via photolithography. In recent years, significant advances have been made in SAW devices thanks to advances in thin film transfer technology. More specifically, lithium tantalate and lithium niobate (LN) on insulator platforms have enabled the implementation of high acoustic velocity substrates [3], [4], [5], allowing for enhanced confinement of the acoustic wave at the surface of the film stack. This comes with benefits for the electromechanical coupling ($K^2$) and quality factor (Q), both key figures of merit for acoustic devices. This has allowed for increased performance of SAW-based devices such as filters [6], [7] and acoustic delay lines (ADL) [8], [9], [10] for radio frequency (RF) signal processing.

For applications within RF communications, wideband acoustic transduction is highly desirable. This wideband response could help enable improvements for devices such as ADL, on-chip acousto-optic modulators [11], [12], [13] and acoustic amplifiers [14], [15]. With a conventional design, the achievable bandwidth limit is largely determined by the $K^2$ of the piezoelectric material and the mode selected. While LN on insulator platforms have helped improve this metric through, there is still an intrinsic cap on the achievable bandwidth utilizing an interdigitated transducer (IDT) or single-phase

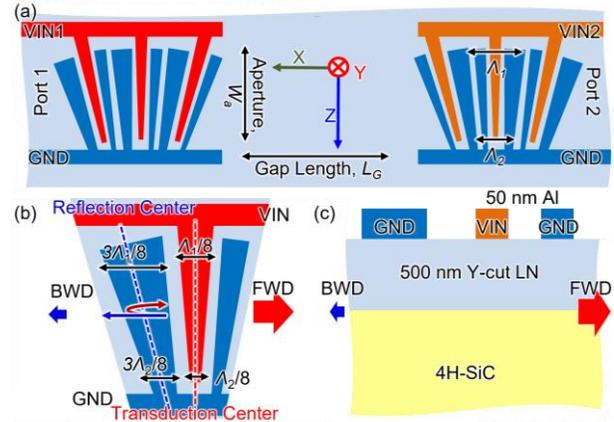

*Figure 1 (a) Top view of the proposed tapered ADL design, with a linear taper from $\Lambda_1$ to $\Lambda_2$ down the length of the transducer. LN crystal orientation is also indicated. (b) Close up depiction of a single transducer, with the corresponding transduction and reflection centers indicated. (c) Cross sectional depiction of the transducer, with layers indicated.*

unidirectional transducer (SPUDT) based design. Conventional SAW transducers utilize alternating electrodes spaced at a distance corresponding to the desired center wavelength of the passband. Thus, a design capable of creating a wideband response may improve the performance of future SAW-based devices.

In this work, we present a tapered single phase unidirectional transducer design for ultra-wideband operation. The design centers around the utilization of a shear horizontal surface acoustic wave (SH-SAW) in Y-cut LN on silicon carbide (SiC), selected for its high electromechanical coupling [16]. The tapered design allows for the excitation of multiple acoustic wavelengths along the aperture of the device, significantly enhancing the total bandwidth. The range of wavelengths excited also spans to overlap the passband of the next order SH-SAW mode, further increasing the total bandwidth. The measured response shows a fractional bandwidth (FBW) of 55% at 2.23 GHz for the measured 50 Ω response. This iteration of the device suffers from poor impedance matching and so performance can be expected to improve to 79% FBW after matching. This has the potential to improve performance of SAW based amplifiers, non-reciprocal elements [17], [18] and acousto-optic devices with further optimization

## DESIGN AND SIMULATION

The proposed design structure is shown schematically in Figure 1 (a)-(c). For this demonstration, a 500 nm Y-cut LN on SiC is selected to optimize for the excitation of the SH-SAW mode along the material X-axis. The SH-SAW mode is chosen for its high electromechanical coupling. Similarly, the SiC substrate is selected for its high acoustic wave velocity, which allows the SH-SAW mode to be tightly confined to the LN layer. The actual transducer design

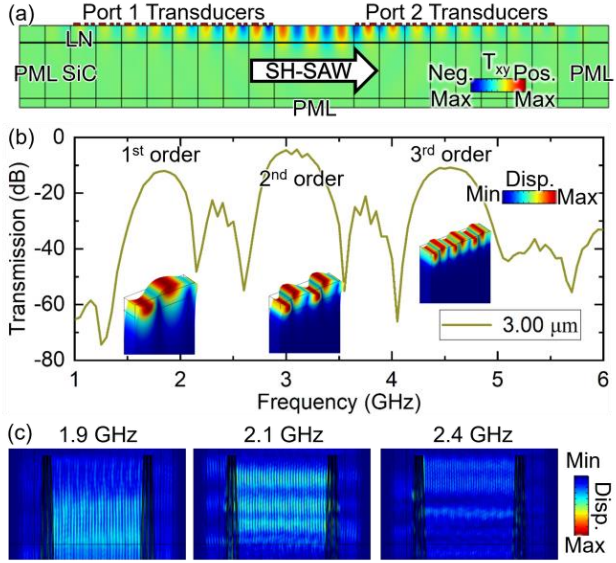

*Figure 2 (a) Cross sectional view of the COMSOL simulation, showing the transmission of the SH-SAW mode from the port 1 transducer, across the gap region, and received by the port 2 transducers. (b) Predicted transducer frequency response for the Λ=3 μm testbed device. The simulation shows the generation of first, second and third order SH-SAW modes from 1.5 GHz to 4.5 GHz.*

consists of a set of 3 SPUDT cells. A unidirectional design is employed to emulate a typical SPUDT, which typically achieves improved insertion loss values while reducing bandwidth. Since this design looks to improve the bandwidth characteristics through the tapering of the IDT, the reduction in bandwidth becomes a secondary concern. The SPUDT relies on the reflection of backward propagating waves through the use of a reflection center transducer finger located $3\Lambda/8$ away from the next IDT electrode. The transducer electrodes are defined with two characteristic wavelengths, $\Lambda_1=6$ μm and $\Lambda_2=3$ μm, where $\Lambda_1$ is the acoustic wavelength at the top of the transducer and $\Lambda_2$ is the wavelength at the bottom. The widths of the individual transducer signal electrodes taper linearly along the length of the finger from $\Lambda_1/8$ to $\Lambda_2/8$. Similarly, the reflection center electrode tapers in width from $3\Lambda_1/8$ to $3\Lambda_2/8$. This reflection tapering maintains the $3\Lambda/8$ phase requirement of an SPUDT at each differential element along the transducer. The tapering is selected to effectively span the transmission spectra of the two characteristic wavelengths, thus ranging from the low-frequency operation of $\Lambda_1$ to the higher frequency operation of $\Lambda_2$. The two tapered transducers are separated with a gap length ($L_g$) of 20 μm, and the aperture (Wa) is chosen to be 200 μm to approximate impedance match the system. The transducers are connected to signal-ground probing pads with a 200 μm pitch.

The expected response from the transducer was then simulated in COMSOL finite element analysis (FEA). A cross-section of the delay line structure is shown in Fig. 2 (a), with the stress in the XY plane plotted at the center of the first SH-SAW passband. From this cross-section, the SH-SAW mode is identified to be largely confined to the LN layer with only minimal stress in the SiC layer. The tight confinement of the acoustic wave to the surface of the LN ensures a greater overlap between the electric field and the stress nodes, allowing for higher coupling in the stack. Next, the single characteristic wavelength of 3 μm is simulated to show the expected frequency response from the highest frequency range of the transducer, with the transmission of the transducer plotted in Fig. 2 (b). Three passbands can be observed in this simulation, with the first corresponding to the fundamental SH-SAW mode and the latter two passbands corresponding to the second and third-order modes, respectively. Since the total response of the transducer is expected to be a superposition of passbands associated with the discrete wavelengths contained along the taper, these higher-order passbands are similarly expected to contribute to a wider total bandwidth of operation. Finally, an iteration of the tapered design is simulated in Fig. 2 (c) to show the transduction at different points along the length of the aperture. The simulation shows a top view of the displacement at three different discrete frequencies, showing the transition from the wider portion of the transducer providing a majority of the excitation at lower frequency to the higher portion exciting the wave at higher frequencies.

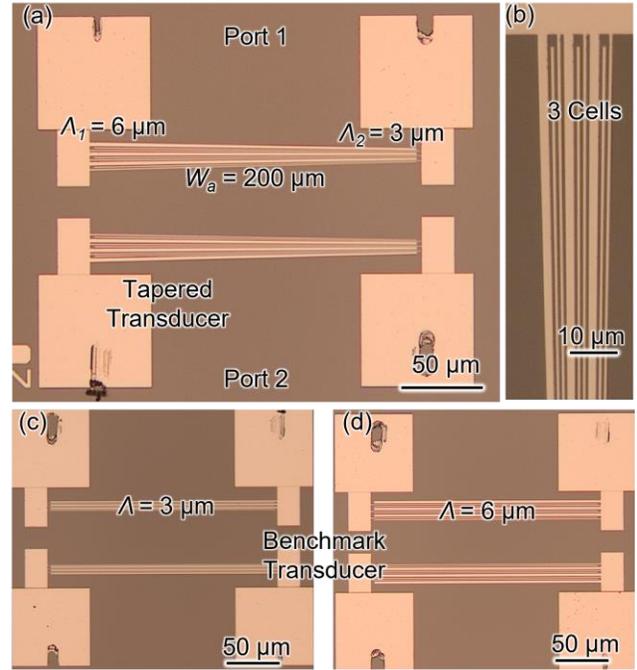

*Figure 3 (a) Fabricated tapered transducers to form an ADL structure. (b) Zoomed in image of the transducer structure. (c)-(d) Testbed ADL platform with an acoustic wavelength of 3 μm and 6 μm respectively.*

## DEVICE FABRICATION AND RESULTS

The fabricated device is shown in Fig. 3 (a) and (b). Additional benchmark transducers, shown in Fig. 3 (c) and (d), are fabricated alongside the tapered transducer design. These benchmark designs allow for the further analysis of performance characteristics, without the additional complexity of the tapered design. The devices were fabricated using a standard fabrication workflow. First, a Y-cut LN on 4H-SiC wafer is obtained from NGK Insulators. The wafer is then diced, and the individual sample is cleaned prior to further fabrication steps. E-beam lithography is then performed to pattern the fine electrode features of the sample. A 50 nm thick aluminum layer is deposited using e-beam evaporation, followed by lift off using acetone and sonication. After a solvent-based clean, a second layer of e-beam lithography, 300nm aluminum metal deposition, and lift-off is then performed to achieve thicker busline and probing pads. The thicker busline features are required to ensure good

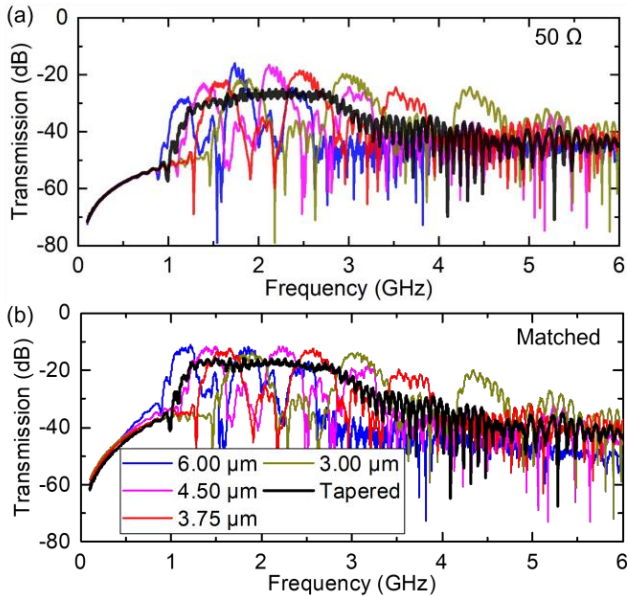

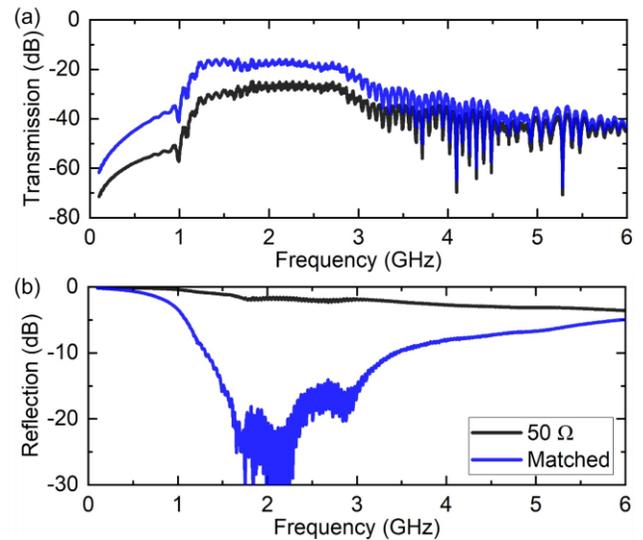

*Figure 4 (a) Measured 50 Ω response of the tapered design, along with 4 benchmark devices. (b) Measured response after conjugate impedance matching with a complex impedance of 235+150j Ω.*

*Figure 5 Plots of transmission (a) and reflection (b) for both the matched and 50 Ω cases. After matching, significant improvements are displayed.*

contact during probing. This also comes with the benefit of reduced electrical resistance on the portion of the device leading up to the transducer electrodes.

The devices are then measured using a Keysight P5028A vector network analyzer (VNA) connected to 200 µm pitch Picoprobe 67A probes from 100 MHz to 6 GHz. The GS probing pads are landed on with one ground finger landed on a dummy pad placed to the side of the device. The Z-axis of the probe is then slowly adjusted until the measurement stabilizes, and good contact is observed through the overlap of the S11 and S22 scattering parameters. The measured frequency response of the tapered transducer design is shown in Fig. 4 (a), along with the response of 4 benchmark transducers with wavelengths ranging from 3 µm to 6 µm. The benchmark transducers show the predicted first through third-order SH-SAW passbands for each of the wavelengths tested. Since longer wavelength transducers have a closer frequency spacing between passbands, all three passbands can be observed to overlap the tapered transducer passband. Similarly, the shorter wavelengths overlap the tapered design for two of the three passbands. The higher order modes of these longer wavelengths become the dominant transduction mechanism beyond the primary 1 to 3 GHz range. This can be used qualitatively to explain the slow roll off of the transducer towards high frequencies. Within the region determined by the first order passbands, the response of the tapered transducer shows a relatively flat passband response from 1.25 GHz to 2.80 GHz. The corresponding FBW is measured to be 55% with a center frequency of 2.23 GHz for the 50Ω measurement. The corresponding insertion loss was measured to be 26 dB at the center frequency. The flat passband and ultrawide FBW are attributed to the integration of the frequency-dependent passbands demonstrated in the benchmark devices. Since the electrodes taper linearly along the length of the aperture, the weighting of frequency components associated with a given location along the transducer is constant and thus yields a largely flat response. Utilizing a curved or stepped transducer design, specific regions could be given a larger weighting to further tune the response.

The measured response was then impedance matched in Keysight ADS, yielding a conjugate match of 235+150j Ω. Plots of the impedance-matched case for tapered and benchmark devices are shown in Fig. 4 (b). After matching, improved metrics could be expected, with an FBW of 79% and IL of 16.5 being demonstrated. Matching was also performed on the benchmark devices, with an IL of ~10 dB being observed for each of the designs. The high IL value for the benchmark devices is likely due to large series electrical resistance associated with the thin 50 nm thick electrodes. This would similarly impact the tapered transducer, resulting in degradation compared to the potential performance metrics achievable. Since the bandwidth requirements are being improved through the use of the tapered design, more transducer unit cells could be used to help mitigate the need for long aperture widths to help mitigate this resistance and achieve closer to a 50Ω match. Thus, through improved impedance matching and design changes, a trade-off can be made between bandwidth and IL depending on the application. In Fig. 5 (a) and (b), isolated plots of the tapered transducer transmission and reflection before and after matching are shown. Prior to matching, only a slight dip in reflection can be observed. However, after matching, the passband primarily associated with the first-order SH-SAW passbands can be clearly identified. This further suggests that the primary limitation of the performance of this design was poor impedance matching. Thus, future iterations may look to increase performance through on and off-chip matching techniques.

## CONCLUSION

In this work, we present a linearly tapered single phase unidirectional transducer for ultra-wideband performance. The design shows a measured FBW of 55% at 2.23 GHz with an IL of 26 dB. This ultrawide performance is enabled through the excitation of SH-SAW ranging in wavelength from 3 µm to 6 µm. Additionally, the second-order SH-SAW mode helps extend the passband further. With improved impedance matching, this performance can be enhanced to 79% FBW and an IL of 16.5 dB. The heightened IL values are likely due to large series resistance values associated with the long and thin electrodes. Thus, improvements in IL can be made by trading off bandwidth with a larger cell count while still benefiting from the wide frequency range of the tapered design. Upon further optimization, this design can be tailored to improve the wideband operation of SAW-based signal processing elements.

## ACKNOWLEDGEMENTS

The authors would like to thank the support of the Sandia LDRD for the funding and Dr. Lisa Hackett and Dr. Matt Eichenfield for the helpful discussions.